\documentclass[aps,prd,superscriptaddress,preprint,tightenlines,nofootinbib]{revtex4}

\usepackage{graphicx} 
\usepackage{dcolumn}  
\usepackage{bm}       

\begin{document}

\preprint{CLNS 08/2047}       
\preprint{CLEO 08-29}         

\title{\boldmath Determination of the $D^{0}\to K^{-}\pi^{+}\pi^{0}$ and 
$D^{0}\to K^{-}\pi^{+}\pi^{+}\pi^{-}$ Coherence Factors and Average Strong-Phase Differences Using Quantum-Correlated 
Measurements}



\author{N.~Lowrey}
\author{S.~Mehrabyan}
\author{M.~Selen}
\author{J.~Wiss}
\affiliation{University of Illinois, Urbana-Champaign, Illinois 61801, USA}
\author{R.~E.~Mitchell}
\author{M.~R.~Shepherd}
\affiliation{Indiana University, Bloomington, Indiana 47405, USA }
\author{D.~Besson}
\affiliation{University of Kansas, Lawrence, Kansas 66045, USA}
\author{T.~K.~Pedlar}
\affiliation{Luther College, Decorah, Iowa 52101, USA}
\author{D.~Cronin-Hennessy}
\author{K.~Y.~Gao}
\author{J.~Hietala}
\author{Y.~Kubota}
\author{T.~Klein}
\author{R.~Poling}
\author{A.~W.~Scott}
\author{P.~Zweber}
\affiliation{University of Minnesota, Minneapolis, Minnesota 55455, USA}
% ...
\collaboration{CLEO Collaboration}
\noaffiliation

\date{\today}
 
\begin{abstract} 
The first measurements of the coherence factors ($R_{K\pi\pi^{0}}$ and $R_{K3\pi}$) and the average strong-phase 
differences ($\delta_{D}^{K\pi\pi^{0}}$ and $\delta_{D}^{K3\pi}$) for $D^{0}\to K^{-}\pi^{+}\pi^{0}$ and $D^{0}\to 
K^{-}\pi^{+}\pi^{+}\pi^{-}$ are presented. These parameters can be used to improve the determination of the unitarity 
triangle angle $\gamma$ in $B^{-}\to DK^{-}$ decays, where $D$ is a $D^{0}$ or $\bar{D}^{0}$ meson decaying to the 
same final state. The measurements are made using quantum-correlated, fully-reconstructed  $D^0\bar{D^0}$ pairs 
produced  in $e^+e^-$ collisions at the $\psi(3770)$ resonance. 
The measured values are:   $R_{K\pi\pi^{0}}=0.84\pm 0.07$, 
$\delta_{D}^{K\pi\pi^{0}}=(227^{+14}_{-17})^{\circ}$, $R_{K3\pi}=0.33^{+0.20}_{-0.23}$, and 
$\delta_{D}^{K3\pi}=(114^{+26}_{-23})^{\circ}$. These results indicate significant coherence in the decay $D^{0}\to 
K^{-}\pi^{+}\pi^{0}$, whereas lower coherence is observed in the decay $D^{0}\to K^{-}\pi^{+}\pi^{+}\pi^{-}$. The 
analysis also results in a small improvement in the knowledge of other $D$-meson parameters, in particular the 
strong-phase difference for $D^{0}\to K^{-}\pi^{+}$, $\delta_{D}^{K\pi}$, and the mixing parameter, $y$.  
\end{abstract}

\maketitle

 This paper presents the first determination of the coherence factors and the average strong-phase differences for 
$D^{0}\to K^{-}\pi^{+}\pi^{0}$ and $D^{0}\to K^{-}\pi^{+}\pi^{+}\pi^{-}$ made using quantum-correlated, 
fully-reconstructed (double-tagged) $D^0\bar{D^0}$ pairs produced  in $e^+e^-$ collisions at the $\psi(3770)$ 
resonance.  
Knowledge of these parameters improves the sensitivity of measurements of the unitarity triangle angle $\gamma$ using 
$B$-meson decays to these $D$-meson final states.  Although $CP$-violation involving $B$-mesons has been clearly 
established experimentally \cite{bib:CPV}, and existing results 
are in good agreement with Standard Model predictions, additional and 
improved measurements are required to overconstrain the 
Cabibbo-Kobayashi-Maskawa (CKM) quark-mixing matrix \cite{bib:CKM} and probe for the effects of 
non-Standard Model physics.  An important ingredient in this program will be a precise determination
of the angle $\gamma$.

 Several methods to determine $\gamma$ using $B^{-}\to DK^{-}$ \cite{bib:conjugation} decays 
have been proposed \cite{bib:GLW,bib:ADS,bib:GGSZ}. Here, $D$ refers to either a $D^{0}$ or 
$\bar{D}^{0}$ meson. All these methods exploit the fact that a $B^{-}$ can decay into 
$D^{0}K^{-}$ and $\bar{D}^{0}K^{-}$ final states via $b\to c\bar{u}s$ and $b\to u\bar{c}s$ 
transitions, respectively. The weak phase between these two transitions is equal to $-\gamma$. Therefore, 
the amplitudes  are related by:
$\mathcal{A}(B^{-}\to \bar{D}^{0}K^{-})/\mathcal{A}(B^{-}\to D^{0}K^{-}) = r_{B}e^{i(\delta_{B}-\gamma)}$,  
where $r_B\sim 0.1$ is the absolute amplitude ratio and $\delta_B$ is the strong-phase difference. 
The two amplitudes interfere with one another if the $D^{0}$ and $\bar{D}^{0}$ decay to the same final 
state, which can lead to direct $CP$-violation between the $B^{-}$ and $B^{+}$ decay rates if $\gamma$ 
is non-zero.
        
The Atwood-Dunietz-Soni (ADS) method \cite{bib:ADS} uses common flavor-specific final states such as $D\to 
K^{-}\pi^{+}$ to determine $\gamma$. The rates are given by:
\begin{equation}
\begin{array}{lcl}
\multicolumn{3}{l}{\Gamma(B^{\mp}\to D(K^{\mp}\pi^{\pm})K^{\mp})} \\
 &\propto & 1 + (r_{B}r_{D}^{K\pi})^{2} +  2r_{B}r_{D}^{K\pi}\cos{(\delta_{B}-\delta_D^{K\pi}\mp \gamma)} 
\label{eqn:adsfavoured}
\end{array}
\end{equation}
and
\begin{equation}
\begin{array}{lcl}
\multicolumn{3}{l}{\Gamma(B^{\mp} \to  D(K^{\pm}\pi^{\mp})K^{\mp})}\\
 &\propto &  (r_{B})^{2} + (r_{D}^{K\pi})^{2} \nonumber + 2r_{B}r_{D}^{K\pi}\cos{(\delta_{B}+\delta_D^{K\pi}\mp 
\gamma)} \label{eqn:adssuppressed}   
\end{array}
\end{equation}
where $r_{D}^{K\pi}$ is the absolute amplitude ratio of the doubly-Cabibbo-suppressed (DCS) decay $D^{0}\to 
K^{+}\pi^{-}$ to the Cabibbo-favored (CF) decay $D^{0}\to K^{-}\pi^{+}$, and  
$\delta_{D}^{K\pi}$ is the strong-phase difference between these two amplitudes, which is defined as:
$\mathcal{A}(D^{0}\to K^{+}\pi^{-})/\mathcal{A}(D^{0}\to K^{-}\pi^{+}) = r_{D}^{K\pi} e^{-i\delta_{D}^{K\pi}}$.
Present measurements give 
$r_D^{K\pi}=0.0579\pm 0.0007$ \cite{bib:HFAG08}, therefore, 
the terms on the righthand side of Eq.~(\ref{eqn:adssuppressed}) are all of the same order, which allows 
significant changes to $\Gamma(B^{\mp}  \to  D(K^{\pm}\pi^{\mp})K^{\mp})$ 
depending on the values of $\gamma$ and the strong phases. The suppressed decays $B^{\mp}  \to  
D(K^{\pm}\pi^{\mp})K^{\mp}$ have not yet been observed \cite{bib:babarads,bib:belleads}. The measurement of 
$\delta_{D}^{K\pi}$ has been made in quantum-correlated $D^{0}\bar{D}^{0}$ decays \cite{bib:tqcaprd} in a similar 
manner to the analysis reported in this paper.


The flavor-specific final states $D\to K^{-}\pi^{+}\pi^{+}\pi^{-}$ ($D\to K^{-}3\pi$) and $D\to K^{-}\pi^{+}\pi^{0}$ 
have significantly larger branching fractions than $D\to K^{-}\pi^{+}$ \cite{bib:PDG08}.
However, for three- or four-body $D$ decay the amplitude ratio and strong-phase difference vary over 
phase space.
For such $D$ decays, for example $D \to K^-\pi^+\pi^0$, Eq.~(\ref{eqn:adssuppressed}) is modified as follows 
\cite{bib:AS}:
\label{eqn:kpipi0adsfavoured}
\begin{equation}
\begin{array}{lcl} 
\multicolumn{3}{l}{\Gamma(B^{\mp}\to D(K^{\pm}\pi^{\mp}\pi^{0})K^{\mp})} \\
 &  \propto & (r_{B})^{2} + (r_{D}^{K\pi\pi^{0}})^{2} \\ 
& & +  2r_{B}r_{D}^{K\pi\pi^{0}}R_{K\pi\pi^{0}}\cos{(\delta_{B}+\delta_D^{K\pi\pi^{0}}\mp \gamma)}
\label{eqn:kpipi0adssuppressed}\; , 
\end{array}
\end{equation}         
where $R_{K\pi\pi^{0}}$, $\delta_{K\pi\pi^{0}}$ and $r_{D}^{K\pi\pi^{0}}$ are defined as:
\begin{eqnarray*}
 R_{K\pi\pi^{0}}e^{-i\delta_{D}^{K\pi\pi^{0}}} & = & 
 \frac{\int 
\mathcal{A}_{K^{-}\pi^{+}\pi^{0}}(\mathbf{x})\mathcal{A}_{K^{+}\pi^{-}\pi^{0}}(\mathbf{x})d\mathbf{x}}{A_{K^{-}\pi^{+}
\pi^{0}}A_{K^{+}\pi^{-}\pi^{0}}} \; \mathrm{and} \\
r_{D}^{K\pi\pi^{0}} & =  &\frac{A_{K^{+}\pi^{-}\pi^{0}}}{A_{K^{-}\pi^{+}\pi^{0}}} \; .\end{eqnarray*}
Here $\mathcal{A}_{K^{\pm}\pi^{\mp}\pi^{0}}(\mathbf{x})$ is the amplitude for $D^{0}\to K^{\pm}\pi^{\mp}\pi^{0}$ at a 
point in multi-body phase space described by parameters $\mathbf{x}$, and $A_{K^{\pm}\pi^{\mp}\pi^{0}}^{2}=\int 
|\mathcal{A}_{K^{\pm}\pi^{\mp}\pi^{0}}(\mathbf{x})|^{2}d\mathbf{x}$. (The expressions for $D \to K^-\pi^+\pi^+\pi^-$ 
take the same form and involve the parameters $r_D^{K3\pi}$, $R_{K3\pi}$ and $\delta_D^{K3\pi}$.)

The parameter $R_{K\pi\pi^{0}}$ is known as the coherence factor and can take any value from zero to one. A small 
value of $R_{K\pi\pi^{0}}$ indicates a lack of coherence between the intermediate states involved in the decay, a 
situation expected
when there are many resonances contributing; a value close to one occurs when the resonances are largely in phase, or
one state dominates. Decays to two-body final states, such as $D^{0}\to K^{-}\pi^{+}$, and to $CP$ eigenstates have a 
coherence factor equal to one. 
Even if the coherence is small the rate described by Eq.~(\ref{eqn:kpipi0adssuppressed}) is still useful, because it
possesses high sensitivity to the parameter $r_B$.

The coherence factors $R_F$ and average strong-phase difference, $\delta_D^F$, where $F={K^{-}\pi^{+}\pi^{0}}$ or 
$K^{-}\pi^{+}\pi^{+}\pi^{-}$, can be determined using double-tagged $D^{0}\bar{D}^{0}$ pairs produced in $e^{+}e^{-}$ 
collisions
at the $\psi(3770)$ resonance. The two mesons are produced in a $C$-odd eigenstate and their decays are 
quantum-correlated.
The rate for the two $D$ mesons to decay to states $F$ and $G$ is given by \cite{bib:AS}:       
\begin{eqnarray}
\Gamma(F|G) & =  & 
\Gamma_0\int\int{|\mathcal{A}_F(\mathbf{x})\mathcal{A}_{\bar{G}}(\mathbf{y})-\mathcal{A}_{\bar{F}}
(\mathbf{x})\mathcal{A}_G(\mathbf
{y})|^{2} d\mathbf{x}}d\mathbf{y} \; \nonumber \\
	   & = &\Gamma_0[A_F^{2}A_{\bar{G}}^{2}+A_{\bar{F}}^2 A_G^2  \nonumber \\
	   & & - 2R_FR_G A_F A_{\bar{F}}A_{G}A_{\bar{G}}\cos{(\delta_D^{G}-\delta_D^{F})} ] \label{eqn:general_FG} \;,  
\end{eqnarray}
where $\mathcal{A}_{F}(\mathbf{x})$ $(\mathcal{A}_{G}(\mathbf{y}))$ and $\mathcal{A}_{\bar{F}}(\mathbf{x})$ 
($\mathcal{A}_{\bar{G}}(\mathbf{y})$) 
are the amplitudes of $D^{0}\to F$ ($D^{0}\to G$) and  $D^{0}\to \bar{F}$ ($D^{0}\to \bar{G}$) at points $\mathbf{x}$ 
($\mathbf{y}$) in phase space, respectively, and $\Gamma_0=\Gamma(\psi(3770)\rightarrow D^{0}\bar{D^{0}})$. From 
Eq.~(\ref{eqn:general_FG}) the following
double-tagged rates arise:
\begin{eqnarray}
  \Gamma(F|CP) & = &\Gamma_0 A_{F}^2 A_{CP}^2[1+(r_D^{F})^2 \nonumber \\
   && - 2\lambda_{\pm}r_D^{F}R_{F}\cos{\delta_D^{F}}] \; , \label{eqn:k3picprate} \\
  \Gamma(F|F)  & = & \Gamma_0 A_{F}^2 A_{\bar{F}}^2 [1-R_{F}^{2}] \; , \label{eqn:k3pilikesign} \\
  \Gamma(F|K^{-}\pi) & = & \Gamma_0 A_{F}^2 A_{K^{-}\pi^{+}}^2[(r_D^{K\pi})^2+(r_D^{F})^2  \nonumber \\ 
                     &   &-  2r_D^{K\pi}r_D^{F}R_{F}\cos(\delta_D^{K\pi}-\delta_D^{F})] \; 
\label{eqn:kpilikesign}
\end{eqnarray}
and
\begin{equation}
\begin{array}{lcl}
\multicolumn{3}{l}{\Gamma(K^{\mp}\pi^{\pm}\pi^{0}|K^{\mp}\pi^{\pm}\pi^{\pm}\pi^{\mp})} \\ 
& = & \Gamma_0 A_{K^{-}\pi^{+}\pi^{0}}^2 A_{K^{-}\pi^{+}\pi^{+}\pi^{-}}^2 [(r_D^{K3\pi})^2 +(r_D^{K\pi\pi^{0}})^2 \\ 
&&- 2r_D^{K3\pi}r_D^{K\pi\pi^{0}}R_{K3\pi}R_{K\pi\pi^{0}}\cos(\delta_D^{K3\pi}-\delta_D^{K\pi\pi^{0}})] \; .
\label{eqn:ninth}
\end{array}
\end{equation}
Here $CP$ denotes a $CP$ eigenstate with eigenvalue $\lambda_{\pm}=\pm 1$.
The final states described by Eqs.~(\ref{eqn:k3pilikesign}), (\ref{eqn:kpilikesign}), and (\ref{eqn:ninth}) are 
referred to as `like-sign' (LS) on account of the charges of the two kaons involved. 
Furthermore, the following relations are noted: 
$\Gamma(F|CP)=\Gamma(\bar{F}|CP)$, $\Gamma(F|F)=\Gamma(\bar{F}|\bar{F})$ and 
$\Gamma(F|K^{-}\pi^{+})=\Gamma(\bar{F}|K^{+}\pi^{-})$; these expressions 
ignore $CP$-violation in $D$ decay, which is well motivated theoretically and by current experimental limits 
\cite{bib:charmrev}. 

To relate the amplitudes in Eqs.~(\ref{eqn:k3picprate}) to (\ref{eqn:ninth}) to branching fractions the 
effects of charm mixing must be included. Charm mixing is commonly characterised by the parameters 
$x=(M_{+}-M_{-})/\Gamma$ and 
$y=(\Gamma_{+}-\Gamma_{-})/2\Gamma$, where $M_{\pm}$ and $\Gamma_{\pm}$ are the 
masses and widths of the $\lambda_{\pm}=\pm 1$ neutral $D$ meson mass eigenstates, respectively, 
and $\Gamma = (\Gamma_{+}+\Gamma_{-})/2$. The relations between amplitudes and branching fractions, following 
Ref.~\cite{bib:zzxing}, are given in Table~\ref{tab:mixing_correction}.

The best constraints on $x$, $y$ and $\delta_{D}^{K\pi}$ come from the combination of several measurements 
\cite{bib:charmrev}. These constraints \cite{bib:HFAG08} are included in the analysis reported here to improve the 
determination of $R_{F}$ and $\delta_{D}^{F}$. However, the analysis is also sensitive to these parameters so results 
are presented without the external constraints as well.

\begin{table}[th]
\caption{Relations between branching fractions, $\mathcal{B}$, and amplitudes including the 
effects of charm mixing. The DCS and $CP$ expressions are quoted to $\mathcal{O}((x/r_D)^2,(y/r_D)^2)$ and 
$\mathcal{O}(y)$, respectively. The corrections due to mixing in the CF amplitude are negligible $(<1\%)$.}
\label{tab:mixing_correction}.
\begin{tabular}{clr} \hline\hline
Mode & \multicolumn{2}{c}{$\mathcal{B}$} \\ \hline
$D^{0}\to CP$ &  $A_{CP}^2 (1 -\lambda_{\pm} y)$ &\\
$D^{0}\to F$ & $A_{F}^{2}$ &\\
$D^{0}\to \bar{F}$ & $A_{\bar{F}}^{2}[1  -  (y / r_D^{F})\, R_{F}\cos{\delta_D^{F}}$ & \\ 
& \multicolumn{2}{c}{$+  (x / r_D^{F})\, R_{F}\sin{\delta_D^{F}} + (y^2 + x^2)/2(r_D^{F})^2]$} \\
$D^{0}\to K^{-}\pi^{+}$ & $A_{K^{-}\pi^{+}}^{2}$ & \\
$D^{0}\to K^{+}\pi^{-}$ & $A_{K^{+}\pi^{-}}^{2}[1 - (y / r_D^{K\pi})\,\cos \delta_D^{K\pi}$ &\\
& \multicolumn{2}{c}{$ + (x / r_D^{K\pi})\,\sin \delta_D^{K\pi} + (y^2 + x^2)/2(r_D^{K\pi})^2]$} \\ \hline\hline
\end{tabular}
\end{table}

 An $818~\mathrm{pb^{-1}}$ data set of $e^{+}e^{-}$ collisions produced by the Cornell Electron Storage Ring (CESR) at 
$E_{\mathrm{cm}}=3.77~\mathrm{GeV}$ and collected with the CLEO-c detector is analysed. The CLEO-c detector is 
described in detail elsewhere \cite{bib:cleoc}. Table~\ref{tab:finalstates} lists the reconstructed $D^{0}$ and 
$\bar{D}^{0}$ final states, with $\pi^{0}\to\gamma\gamma$, $K^{0}_{S}\to\pi^{+}\pi^{-}$, 
$\omega\to\pi^{+}\pi^{-}\pi^{0}$, $\phi\to K^{+}K^{-}$, $\eta\to\gamma\gamma$, $\eta\to\pi^{+}\pi^{-}\pi^{0}$ and 
$\eta^{\prime}\to\eta(\gamma\gamma)\pi^{+}\pi^{-}$. When required in the analysis, reconstruction efficiencies are 
calculated from simulated samples of signal $D$ decays. Backgrounds from other $D\bar{D}$ decays are estimated from a 
simulated sample of generic $D\bar{D}$ decays.

\begin{table}[th]
\begin{center}
\caption{$D$ final states reconstructed in this analysis.}
\label{tab:finalstates}
\begin{tabular}{cc}\hline\hline
Type & Final states \\ \hline
Flavored & $K^{\mp}\pi^{\pm}$, $K^{\mp}\pi^{\pm}\pi^{\pm}\pi^{\mp}$, $K^{\mp}\pi^{\pm}\pi^{0}$ \\
$CP$-even & $K^{+}K^{-}$, $\pi^{+}\pi^{-}$, $K^{0}_{S}\pi^{0}\pi^{0}$, $K^{0}_{L}\pi^{0}$, $K^{0}_{L}\omega$ \\
$CP$-odd  & $K^{0}_{S}\pi^{0}$, $K^{0}_{S}\omega$, $K^{0}_{S}\phi$, $K^{0}_{S}\eta$, $K^{0}_{S}\eta^{\prime}$ \\ 
\hline   
\end{tabular}
\end{center}
\end{table}

\begin{figure}
\begin{tabular}{cc}
\includegraphics*[width=0.495\columnwidth]{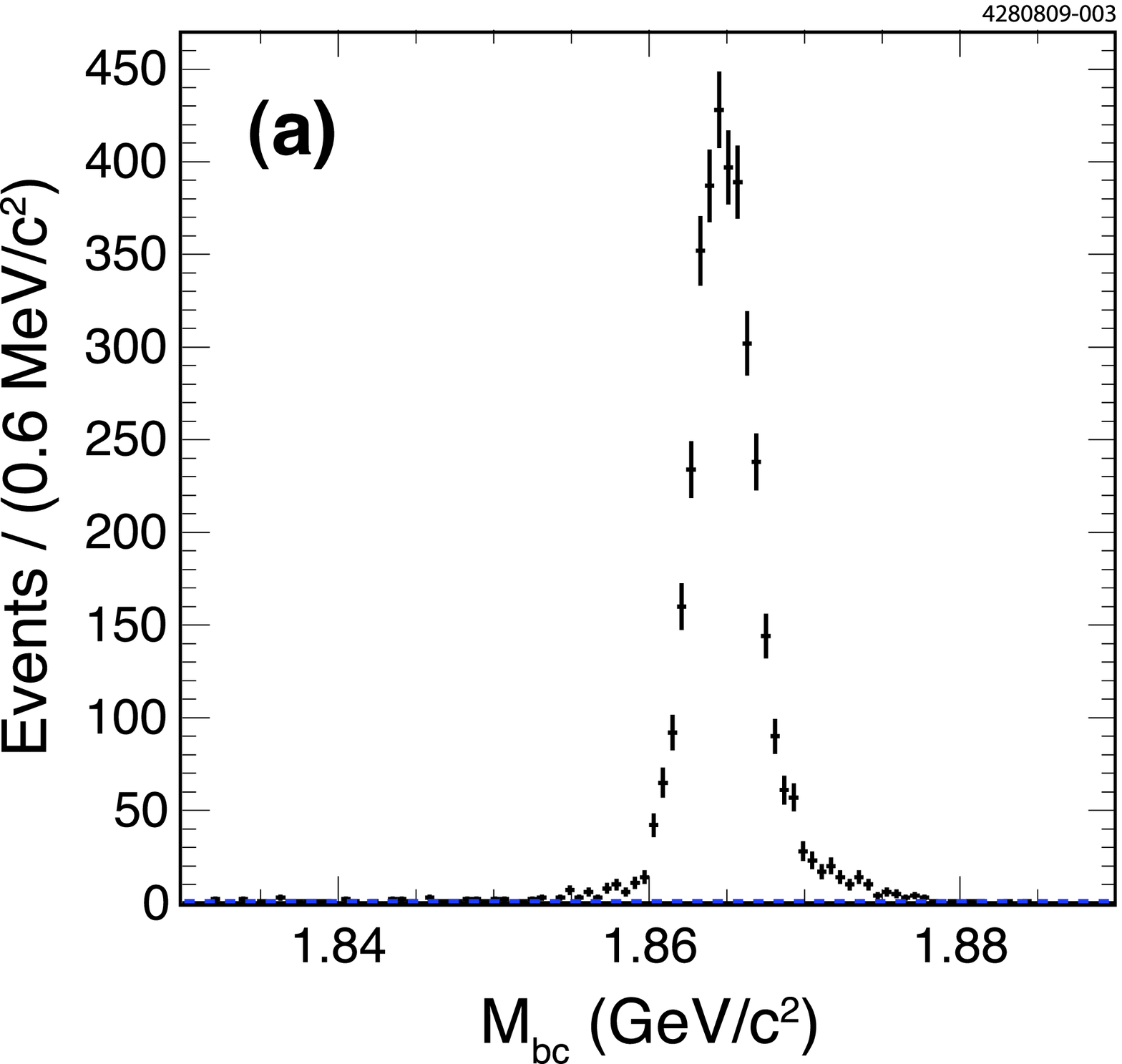} & 
\includegraphics*[width=0.495\columnwidth]{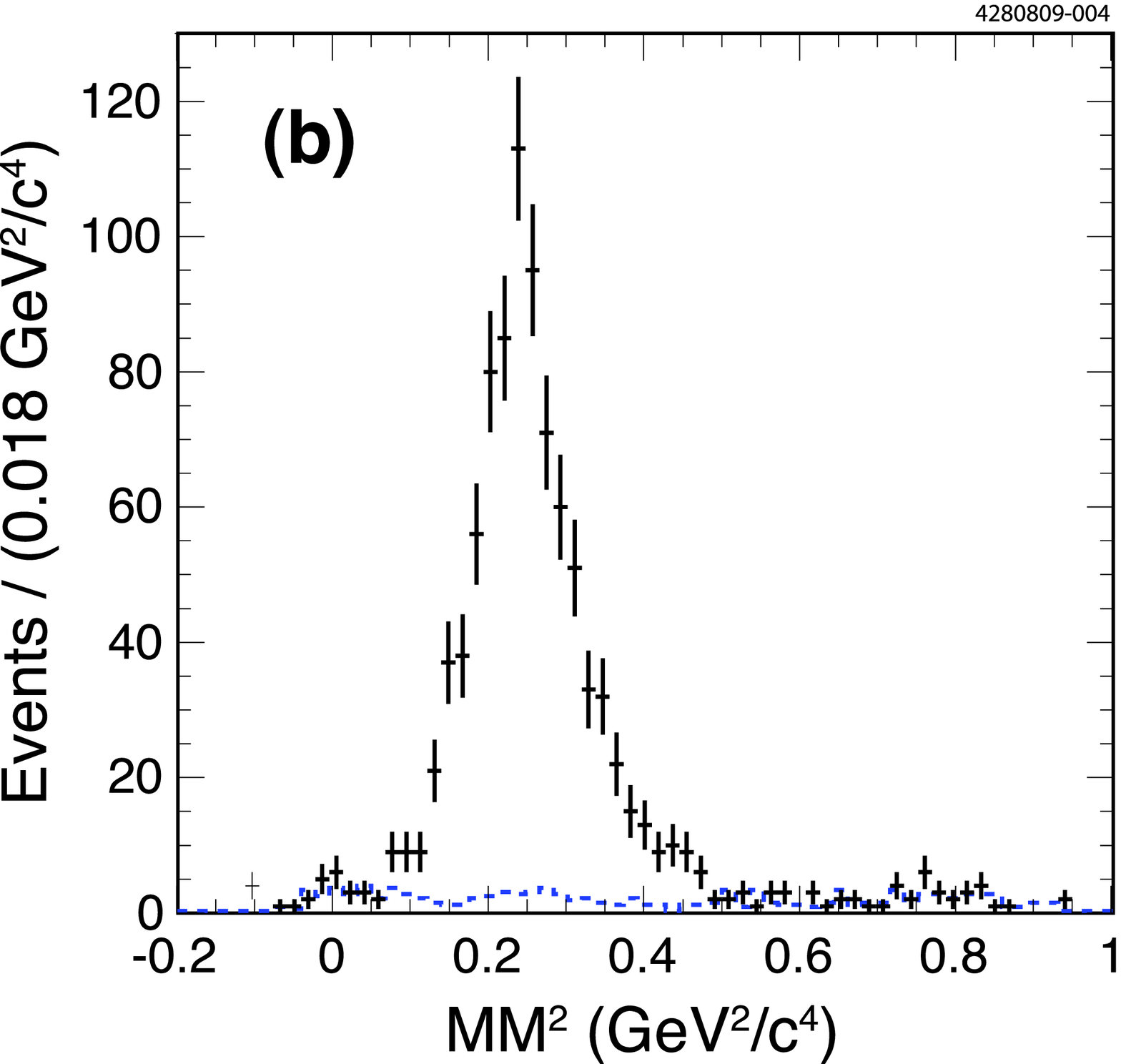}
\end{tabular}
\caption{Distributions of (a) $M_{bc}$ for $D\to K^{-}\pi^{+}\pi^{0}$ candidates tagged by $D\to K^{0}_{S}\pi^{0}$ and 
(b) missing-mass squared for $D\to K^{0}_{L}\pi^{0}$ tagged $D\to K3\pi$ candidates for data (points) and expected 
background (dotted line).} \label{fig:selvar}  
\end{figure}

The $\pi_{0}$, $K^{0}_{S}$, $\omega$ and $\eta\to\gamma\gamma$ reconstruction is identical to that used in Ref. 
\cite{bib:tqcaprd}. Candidates for $\eta\to\pi^{+}\pi^{-}\pi^{0}$, $\eta^{\prime}$, and $\phi$ mesons are considered 
if their masses  are within the intervals  [506,~590]~MeV/$c^{2}$, [950,~964]~MeV/$c^{2}$, and  
[1009,~1033]~MeV/$c^{2}$, respectively.
Final states that do not contain a $K^{0}_{L}$ meson are fully reconstructed via two kinematic variables: the 
beam-constrained candidate mass, $M_{bc}\equiv\sqrt{E_{\mathrm{cm}}^{2}/(4c^{4})-\mathbf{p}_{D}^{2}/c^{2}}$, where 
$\mathbf{p}_{D}$ is the $D$ candidate momentum, and $\Delta E\equiv E_{D}-E_{\mathrm{cm}}/2$, where $E_{D}$ is the sum 
of the $D$ daughter candidate energies. The double-tagged yield is determined from counting events in signal and 
sideband regions of $M_{bc}$. Fig.~\ref{fig:selvar}~(a) shows the distribution of $M_{bc}$ for $D\to 
K^{-}\pi^{+}\pi^{0}$ candidates tagged by $D\to K_S^{0}\pi^{0}$ decays for data and simulated background events.  The 
selection and yield determination procedures are similar to those presented in Ref.~\cite{bib:tqcaprd}. For modes that 
were not considered in Ref.~\cite{bib:tqcaprd} the values of the $\Delta E$ criteria are identical to those used in  
Ref.~\cite{bib:dhadprd}. In addition, to suppress background from $D^{0}\to K^{0}_{S}K^{\pm}\pi^{\mp}$ to $D^{0}\to 
K^{\pm}\pi^{\mp}\pi^{\mp}\pi^{\pm}$, requirements are placed on the $\pi^{+}\pi^{-}$ pairs to be consistent with 
originating from the $e^{+}e^{-}$ collision point. Furthermore, in events where  $K^{-}3\pi$ or $K^{-}\pi^{+}\pi^{0}$ 
are tagged by $K^{\pm}\pi^{\mp}$, at least one of the daughters of the two-body decay is required to be in the 
acceptance of the Ring Imaging Cherenkov detector; this criterion suppresses events where $K^{-}\pi^{+}$ is 
misidentified as $K^{+}\pi^{-}$. 

To identify $CP$-tags containing a single $K^0_L$ meson, we compute the missing-mass recoiling against the signal $D$ 
candidate and the sister particles in the assumed tag decay, and select events consistent with the mass of the $K^0_L$ 
meson squared \cite{bib:K0Lprl}. Yields are extracted from the signal and sideband regions of the missing-mass 
distribution. Fig.~\ref{fig:selvar}~(b) is the distribution of missing-mass squared for $D\to K3\pi$ candidates tagged 
by $D\to K^{0}_L\pi^{0}$ decays for data and simulated background.     

 Significant peaking backgrounds arise in a few modes: non-resonant decays to $\pi^{+}\pi^{-}\pi^{0}$ for modes 
reconstructed including an $\omega$ or $\eta\to\pi^{+}\pi^{-}\pi^{0}$,  $D\to K^{0}_{S}(\pi^{0}\pi^{0})X$ 
misidentified as $D\to K^{0}_{L}X$ decays, and $D\to K^{0}_{S}(\pi^{+}\pi^{-})K^{-}\pi^{+}$ to $D\to K^{-}3\pi$. 
However, these backgrounds are all smaller than the statistical uncertainty on the yields. The peaking background 
yields are estimated from a simulated sample with a size equivalent to approximately 3.3 times the data sample; the 
uncertainty on the peaking background yield is that due to the statistics of this sample. This uncertainty is added in 
quadrature to that on the combinatoric background subtracted signal yields. There is a further peaking background of 
$D\to K^{+}\pi^{-}$ decays misidentified as $D\to K^{-}\pi^{+}$ for the like-sign $K^{-}3\pi$ or $K^{-}\pi^{+}\pi^{0}$ 
tagged by $K^{\pm}\pi^{\mp}$, which is also estimated from simulated sample. However, this contamination is treated as 
a separate source of systematic uncertainty because it is the dominant source for some measurements. The measured 
event yields after background subtraction are given in Table~\ref{tab:yields}. 
\begin{table}[tb]
\begin{center}
\caption{Measured double-tagged signal yields.} \label{tab:yields}
 \begin{tabular}{cccc} \hline\hline
Mode                     & $K^{\pm}\pi^{\mp}\pi^{\mp}\pi^{\pm}$ & $K^{\pm}\pi^{\mp}\pi^{0}$ & $K^{\pm}\pi^{\mp}$ \\ 
\hline 
$K^{\mp}\pi^{\pm}\pi^{\pm}\pi^{\mp}$  &   $4,044\pm 64$         &    --                     &      --            \\
$K^{\pm}\pi^{\mp}\pi^{\mp}\pi^{\pm}$  &   $29.1\pm 5.9$         &    --                     &      --            \\
$K^{\mp}\pi^{\pm}\pi^{0}$             &   $9,594\pm 99$         &    $7,342\pm 87$          &      --            \\
$K^{\pm}\pi^{\mp}\pi^{0}$             &   $63.6\pm 8.8$         &    $12.5\pm 4.1$          &      --            \\
$K^{\mp}\pi^{\pm}$                    &   $5,206\pm 72$         &    $7,155\pm 85$          &      --            \\
$K^{\pm}\pi^{\mp}$                    &   $35.6\pm 6.2$         &    $7.3\pm 3.3$          &      --            \\ 
\hline
$K^{+}K^{-}$                          &   $536\pm 23$           &    $764\pm 28$            &      --            \\
$\pi^{+}\pi^{-}$                      &   $246\pm 16$           &    $336\pm 18$            &      --            \\
$K^{0}_{S}\pi^{0}\pi^{0}$             &   $283\pm 18$           &    $406\pm 21$            &   $221\pm 15$      \\ 
$K^{0}_{L}\pi^{0}$                    &   $827\pm 30$           &    $1,236\pm 38$          &   $689\pm 28$      \\ 
$K^{0}_{L}\omega$                     &   $296\pm 18$           &    $449\pm 22$            &   $251\pm 17$      \\ 
\hline
$K^{0}_{S}\pi^{0}$                    &   $705\pm 27$           &    $891\pm 30$            &   $473\pm 22$      \\ 
$K^{0}_{S}\omega$                     &   $319\pm 19$           &    $389\pm 21$            &   $183\pm 14$      \\ 
$K^{0}_{S}\phi$                       &   $53.0\pm 7.5$         &    $90.9\pm 9.9$          &   $42.8\pm 6.9$    \\ 
$K^{0}_{S}\eta(\gamma\gamma)$         &   $128\pm 12$           &    $116\pm 11$            &   $65.5\pm 8.3$    \\ 
$K^{0}_{S}\eta(\pi^{+}\pi^{-}\pi^{0})$&   $35.9\pm 6.5$         &    $36.3\pm 7.2$          &   $27.2\pm 5.4$    \\ 
$K^{0}_{S}\eta^{\prime}$              &   $35.7\pm 6.0$         &    $60.6\pm 7.8$          &   $30.0\pm 5.5$    \\ 
\hline\hline 
\end{tabular}
\end{center}
\end{table}

The results of the analysis are presented in terms of the observables $\rho^{F}_{CP\pm}$, $\rho^{F}_{LS}$, 
$\rho^{F}_{K\pi,LS}$ and $\rho^{K\pi\pi^{0}}_{K3\pi,LS}$, which are the ratios of the measured values 
of $\Gamma(F|CP)$, $\Gamma(F|F)$, $\Gamma(F|K^{-}\pi^{+})$ and 
$\Gamma(K^{\mp}\pi^{\pm}\pi^{0}|K^{\mp}\pi^{\pm}\pi^{\pm}\pi^{\mp})$ to the expected rates, on the assumption that the 
two $D$ mesons decay in an uncorrelated
fashion or have zero coherence. 
Therefore significant deviation of any of the $\rho$ parameters from a value of one
can only come about through the quantum-correlated nature of $D\bar{D}$ production
at the $\psi(3770)$ and a non-zero coherence in the $D$ decay.   
The $\rho$ observables are related to the
background and efficiency corrected signal yields, $S$, as follows:
\begin{eqnarray}
\rho^{F}_{LS}  & =  & \frac{S(F | F)+S(\bar{F}|\bar{F})}
{2 N_{D^{0}\bar{D^{0}}} \mathcal{B}(D^{0}\rightarrow 
F)\mathcal{B}(D^{0}\rightarrow \bar{F})} \, , \label{eq:rhols} \\
\rho^{F}_{K\pi, LS} &= &  [S(F| K^{-}\pi^{+})+S(\bar{F}|K^{+}\pi^{-})]/ \nonumber \\
&& 2 N_{D^{0}\bar{D^{0}}}  [ \mathcal{B}(D^{0}\rightarrow F) 
\mathcal{B}(D^{0}\rightarrow K^{+}\pi^{-}) \, + \, \nonumber \\ 
& & \hspace*{1.3cm} \mathcal{B}(D^{0}\rightarrow \bar{F}) 
\mathcal{B}(D^{0}\rightarrow K^{-}\pi^{+}) ] \; , \label{eq:rhokpls} \\
\rho^{F}_{CP\pm} & = &   [S(F|CP)+ S(\bar{F}|CP)]/\nonumber \\
& &  2 N_{D^{0}\bar{D^{0}}} \mathcal{B}(D^{0}\rightarrow CP) \nonumber \\
& & \hspace*{1.3cm} [ \mathcal{B}(D^{0}\rightarrow F) + \mathcal{B}(D^{0}\rightarrow\bar{F}) ] 
\, , \label{eq:rhocp}
\end{eqnarray}
and 
\begin{equation}
\begin{array}{lcl}
\multicolumn{3}{l}{\rho^{K\pi\pi^{0}}_{K3\pi, LS}} \\
 & = &  [S(K^{-}\pi^{+}\pi^{0}| K^{-}3\pi)+S(K^{+}\pi^{-}\pi^{0}|K^{+}3\pi)]/
\\
 && 2 N_{D^{0}\bar{D^{0}}}  [ \mathcal{B}(D^{0}\rightarrow K^{-}\pi^{+}\pi^{0}) 
 \mathcal{B}(D^{0}\rightarrow K^{+}3\pi) \, + \, 
 \nonumber \\
 && \hspace*{1.3cm} \mathcal{B}(D^{0}\rightarrow K^{+}\pi^{-}\pi^{0}) 
\mathcal{B}(D^{0}\rightarrow K^{-}3\pi) ],
\end{array} 
\end{equation}
where $N_{D^0\bar{D^0}}$ is the total number of $\psi(3770) \to D^0\bar{D^0}$ events.

In the extraction of each like-sign observable, the product of $N_{D^{0}\bar{D^{0}}}$ and the
reconstruction efficiency is determined from the background-subtracted yield in the corresponding opposite-sign 
samples, 
taking the values~of the branching fractions reported in Ref.~\cite{bib:PDG08}. 
For example, in the case of $\rho^{F}_{LS}$, the observable is given by
\begin{equation}
\rho^{F}_{LS}  =  \frac{N(F | F)+N(\bar{F}|\bar{F})}{2N(F|\bar{F})}
\frac{\mathcal{B}(D^{0}\rightarrow F)}{\mathcal{B}(D^{0}\rightarrow \bar{F})} \, , \label{eq:rholsnorm} \\
\end{equation}
where $N$ are the background-subtracted yields without any efficiency corrections applied.
For the majority of the $CP$ double-tags an alternative normalization procedure is exploited, whereby
knowledge of $N_{D^0\bar{D^0}}$, the reconstruction efficiency and the branching ratio of the $CP$ mode,
which in many cases is poorly known,  is accommodated by a comparison with double-tag events 
involving the $CP$-tag against $D \to K^-\pi^+$ decays.  The good knowledge of 
$\delta_D^{K\pi}$~\cite{bib:tqcaprd,bib:HFAG08,bib:phase_conv} 
allows the contribution from quantum-correlations in these normalization events to be accounted for. 
Small corrections are applied related 
to the differing environment in which the tag is reconstructed in $K^-\pi^+$, $K^-\pi^+\pi^{0}$ and 
$K^-\pi^+\pi^-\pi^+$ events. In the case of the tags $K^+K^-$ and $\pi^+\pi^-$ the branching ratios are known well 
enough to use the values directly from Ref.~\cite{bib:PDG08}, together with measurements of the reconstruction 
efficiency and $S(F|K^{+}\pi^{-})$.


Table~\ref{tab:rho_observables} shows the measured value of each observable.  In the case of $\rho^F_{CP\pm}$
the results from the individual $CP$-tags are found to be consistent and are 
therefore combined into mean values for $CP$-even and $CP$-odd, taking full account of the correlations among the 
assigned systematic uncertainties.
The most important systematic uncertainties are those arising from the finite size of the 
$K^-\pi^+$ vs. $CP$ double-tag samples (0.018), residual corrections associated with this normalization procedure 
(0.008),
and knowledge of the CF $D^0 \to K^-\pi^+\pi^-\pi^+$ and $D^0 \to K^-\pi^+\pi^0$ branching ratios (0.010).
For the $\rho^{K3\pi}_{LS}$ ($\rho^{K\pi\pi^{0}}_{LS}$) observable the dominant uncertainty of 0.082 (0.021) comes 
from the knowledge of the DCS branching ratio; this is also a significant component for $\rho^{K3\pi}_{K\pi,LS}$ 
($\rho^{K\pi\pi^{0}}_{K\pi,LS}$), where the uncertainty is 0.034 (0.003),
with further important contributions arising from the knowledge of the $D^0 \to K^+\pi^-$ branching ratio of 0.024 
(0.005) and the rate of misidentification of $D^{0}\to K^{-}\pi^{+}$ as $D^{0}\to K^{+}\pi^{-}$ of 0.016 (0.026). For 
$\rho^{K\pi\pi^0}_{K3\pi,LS}$ the largest uncertainty is 0.065 from the DCS branching fractions.
For all observables uncertainties are also assigned to account for non-uniform acceptance across phase-space;
this uncertainty is only found to be significant for $\rho^{K3\pi}_{LS}$, $\rho^{K3\pi}_{K\pi,LS}$ and 
$\rho_{K3\pi,LS}^{K\pi\pi^{0}}$ where it is 0.051, 0.040 and 0.037, respectively. 
The results in Table~\ref{tab:rho_observables} suggest significant coherence in the $D \to K^-\pi^+\pi^0$ decay,
but much less so in the case of $D \to K^-\pi^+\pi^-\pi^+$.

\begin{table}[th]
\caption{Measured $\rho$ observables, where the first uncertainty is statistical and the second systematic.}
\label{tab:rho_observables}.

\begin{tabular}{crcrcr} \hline\hline
Observable & \multicolumn{5}{c}{Measured Value} \\ \hline
$\rho^{K3\pi}_{CP+}$         & $1.077$ & $\pm$ & $0.024$ & $\pm$ & $0.029$ \\
$\rho^{K3\pi}_{CP-}$         & $0.933$ & $\pm$ & $0.027$ & $\pm$ & $0.046$ \\
$\rho^{K3\pi}_{LS}$          & $1.112$ & $\pm$ & $0.226$ & $\pm$ & $0.102$ \\
$\rho^{K3\pi}_{K\pi, LS}$    & $0.971$ & $\pm$ & $0.169$ & $\pm$ & $0.062$ \\ \hline

$\rho^{K\pi\pi^0}_{CP+}$     & $1.073$ & $\pm$ & $0.020$ & $\pm$ & $0.035$ \\
$\rho^{K\pi\pi^0}_{CP-}$     & $0.868$ & $\pm$ & $0.023$ & $\pm$ & $0.049$ \\ 
$\rho^{K\pi\pi^0}_{LS}$      & $0.388$ & $\pm$ & $0.127$ & $\pm$ & $0.026$ \\
$\rho^{K\pi\pi^0}_{K\pi,LS}$ & $0.170$ & $\pm$ & $0.072$ & $\pm$ & $0.027$ \\ \hline

$\rho^{K\pi\pi^0}_{K3\pi,LS}$& $1.221$ & $\pm$ & $0.169$ & $\pm$ & $0.080$ \\ \hline \hline
\end{tabular}
\end{table}


The relationships between the like-sign kaon observables and the physics parameters are given by:
\begin{eqnarray}
 \rho^{F}_{LS} & = & \frac{1-R_{F}^2}
{1+\frac{(x^2+y^2)}{2(r_{D}^{F})^2}-\frac{R_{F}}{r_{D}^{F}}(y\cos{\delta_D^{F}}-x\sin{\delta_D^{F}})} \; ,
\label{eqn:coherence_relationK3piLS} \\
\rho^{F}_{K\pi, LS}  & =  &
\frac{[1+(\frac{r^{F}}{r^{K\pi}})^2-2\frac{r^{F}}{r^{K\pi}}R_{F}\cos{(\delta_D^{K\pi}-\delta_D^{F})}]B^{F}_{K\pi, 
LS}}{1+\frac{(x^2+y^2)}{2(r_{D}^{K\pi})^2}-\frac{1}{r_{D}^{K\pi}}(y\cos{\delta_D^{K\pi}}-x\sin{\delta_D^{K\pi}})} \; , 
\nonumber \\
& & \label{eqn:coherence_relationKpiLS}
\end{eqnarray}
and
\begin{equation}
\begin{array}{l}
\rho^{K\pi\pi^{0}}_{K3\pi, LS}=  \\
\frac{[1+(\frac{r^{K\pi\pi^{0}}}{r^{K3\pi}})^2-2\frac{r^{K\pi\pi^{0}}}
{r^{K3\pi}}R_{K\pi\pi^{0}}R_{K3\pi}\cos{(\delta_D^{K\pi\pi^{0}}-\delta_{D}^{K3\pi})]B^{K\pi\pi^{0}}_{K3\pi, LS}}}{
1+\frac{(x^2+y^2)}{2(r_{D}^{K3\pi})^2}-\frac{R_{K3\pi}}{r_{D}^{K3\pi}}(y\cos{\delta_D^{K3\pi}}-x\sin{\delta_D^{K
3\pi}})} \;,
\end{array}
\end{equation}
where $B^{F}_{K\pi,LS}=\mathcal{B}(D^{0}\to F)\mathcal{B}(D^{0}\to K^{+}\pi^{-})/(\mathcal{B}(D^{0}\to 
F)\mathcal{B}(D^{0}\to K^{+}\pi^{-})+\mathcal{B}(D^{0}\to \bar{F})\mathcal{B}(D^{0}\to K^{-}\pi^{+}))$ and 
$B^{K\pi\pi^{0}}_{K3\pi, LS}=\mathcal{B}(D^{0}\to K^{-}\pi^{+}\pi^{0})\mathcal{B}(D^{0}\to 
K^{+}3\pi)/(\mathcal{B}(D^{0}\to K^{-}\pi^{+}\pi^{0} )\mathcal{B}(D^{0}\to K^{+}3\pi)+\mathcal{B}(D^{0}\to 
K^{+}\pi^{-}\pi^{0})\mathcal{B}(D^{0}\to K^{-}3\pi))$. 

In making use of the $\rho^{F}_{CP\pm}$ observables it is convenient to define the 
$CP$-invariant observable, $\Delta^{F}_{CP}$:
\begin{equation}
\Delta^{F}_{CP} \equiv \lambda_{\pm} (\rho^{F}_{CP\pm} - 1 ) =  y -2 r_{D}^{F}R_{F}\cos{\delta_D^{F}} \; .
\label{eq:deltacp2a}
\end{equation}
Some anticorrelated systematic uncertainties on $\rho^{F}_{CP+}$ and $\rho^{F}_{CP-}$ cancel when computing
$\Delta^{F}_{CP}$. It is found that $\Delta^{K3\pi}_{CP} = 0.077 \pm 0.018 \pm 0.022$ and 
$\Delta^{K\pi\pi^0}_{CP}= 0.097 \pm 0.015 \pm 0.023$, with $\chi^2/d.o.f$ values of
$7.3/10$ and $5.7/10$, respectively.

The values of $R_{K\pi\pi^{0}}$, $R_{K3\pi}$, $\delta_{D}^{K\pi\pi^{0}}$ and 
$\delta_{D}^{K3\pi}$ are obtained by a $\chi^{2}$ fit to $\rho_{LS}^{F}$, $\rho_{K\pi,LS}^{F}$, 
$\rho_{K\pi\pi^{0},LS}^{K3\pi}$ and $\Delta_{CP}^{F}$. In addition, the 
fit has $x$, $y$, $\delta_{D}$ and the CF and DCS branching fractions for $D\to 
K^{-}\pi^{+}$, $K^{-}\pi^{+}\pi^{0}$ and $K^{-}3\pi$ as free parameters. The values 
of the $D$-mixing parameters and branching fractions are constrained to those reported in 
Refs.~\cite{bib:HFAG08} and \cite{bib:PDG08}, respectively; this procedure is referred to as the mixing-constrained 
fit. The values of the constraints are given in Table~\ref{tab:results}.  Correlations amongst all free parameters are 
accounted for. The results of the mixing-constrained fit are given in Tab.~\ref{tab:results}. The best fit values of 
the coherence factors and average 
strong-phase differences are: $R_{K\pi\pi^{0}}=0.84\pm 0.07$, 
$\delta_{D}^{K\pi\pi^{0}}=(227^{+14}_{-17})^{\circ}$, $R_{K3\pi}=0.33^{+0.20}_{-0.23}$, and 
$\delta_{D}^{K3\pi}=(114^{+26}_{-23})^{\circ}$. There are also small improvements in the precision of 
$\delta_{D}^{K\pi}$, $y$ and the DCS and CF $D\to K^{-}3\pi$ branching fractions compared to the external constraints. 
The uncertainties are those arising from the statistical and systematic uncertainties on the observables. The 
$\chi^{2}/d.o.f$ for the mixing-constrained fit is 7.3/3. The correlations amongst the parameters are presented in 
Ref.~\cite{bib:correlations}.

\begin{table*}[htb]
\begin{center}
\caption{Results of the mixing-constained and unconstrained fits to the observables. Values of external constraints 
are listed. The uncertainties are those arising from the statistical and systematic uncertainties on the 
observables.}\label{tab:results}
\begin{tabular}{lccc} \hline\hline
Parameter & Mixing constrained & Mixing unconstrained & External input \\
\hline 
$R_{K\pi\pi^{0}}$                                           
& $0.84\pm 0.07$         & $0.78^{+0.11}_{-0.25}$ & -- \\
$\delta_{D}^{K\pi\pi^{0}}~(^{\circ})$                       
& $227^{+14}_{-17}$      & $239^{+32}_{-28}$      & -- \\
$R_{K3\pi}$                                                 
& $0.33^{+0.26}_{-0.23}$ & $0.36^{+0.24}_{-0.30}$ & -- \\
$\delta_{D}^{K3\pi}~(^{\circ})$                             
& $114^{+26}_{-23}$      & $118^{+62}_{-53}$      & -- \\
$x$ (\%)                                                    
& $0.96\pm 0.25$         & $-0.8^{+2.9}_{-2.5}$     &  $1.00\pm 0.25$\\
$y$ (\%)                                                    
& $0.81\pm 0.16$         & $0.7^{+2.4}_{-2.7}$      &  $0.76\pm 0.18$\\
$\delta_{D}^{K\pi}$                                         
& $-151.5^{+9.6}_{-9.5}$ & $-130^{+38}_{-28}$     &  $-157.5^{+10.4}_{-11.0}$ \\ 
$\mathcal{B}(D^{0}\to K^{-}\pi^{+})$ (\%)                   
& $3.89\pm 0.05$         &     $3.89\pm 0.05$     &  $3.89\pm 0.05$\\
$\mathcal{B}(D^{0}\to K^{+}\pi^{-})~(10^{-4})$              
& $1.47\pm 0.07$         & $1.47\pm 0.07$         &  $1.47\pm 0.07$\\
$\mathcal{B}(D^{0}\to K^{-}\pi^{+}\pi^{0})$ (\%)            
& $13.8\pm 0.5$        & $13.8\pm 0.5$            &  $13.9\pm 0.5$\\
$\mathcal{B}(D^{0}\to K^{+}\pi^{-}\pi^{0})~(10^{-4})$       
& $3.05\pm 0.17$         &  $3.05\pm 0.17$        &  $3.05\pm 0.17$\\
$\mathcal{B}(D^{0}\to K^{-}\pi^{+}\pi^{+}\pi^{-})$ (\%)     
& $7.96\pm 0.19$         & $8.03\pm 0.19$         &  $8.10\pm 0.20$\\
$\mathcal{B}(D^{0}\to K^{+}\pi^{-}\pi^{-}\pi^{+})~(10^{-4})$ 
& $2.65\pm 0.19$         & $2.63\pm 0.19$         &  $2.62\pm 0.20$\\ \hline\hline 
\end{tabular}
\end{center}
\end{table*}

\begin{figure}
\begin{tabular}{cc}
\includegraphics*[width=0.495\columnwidth]{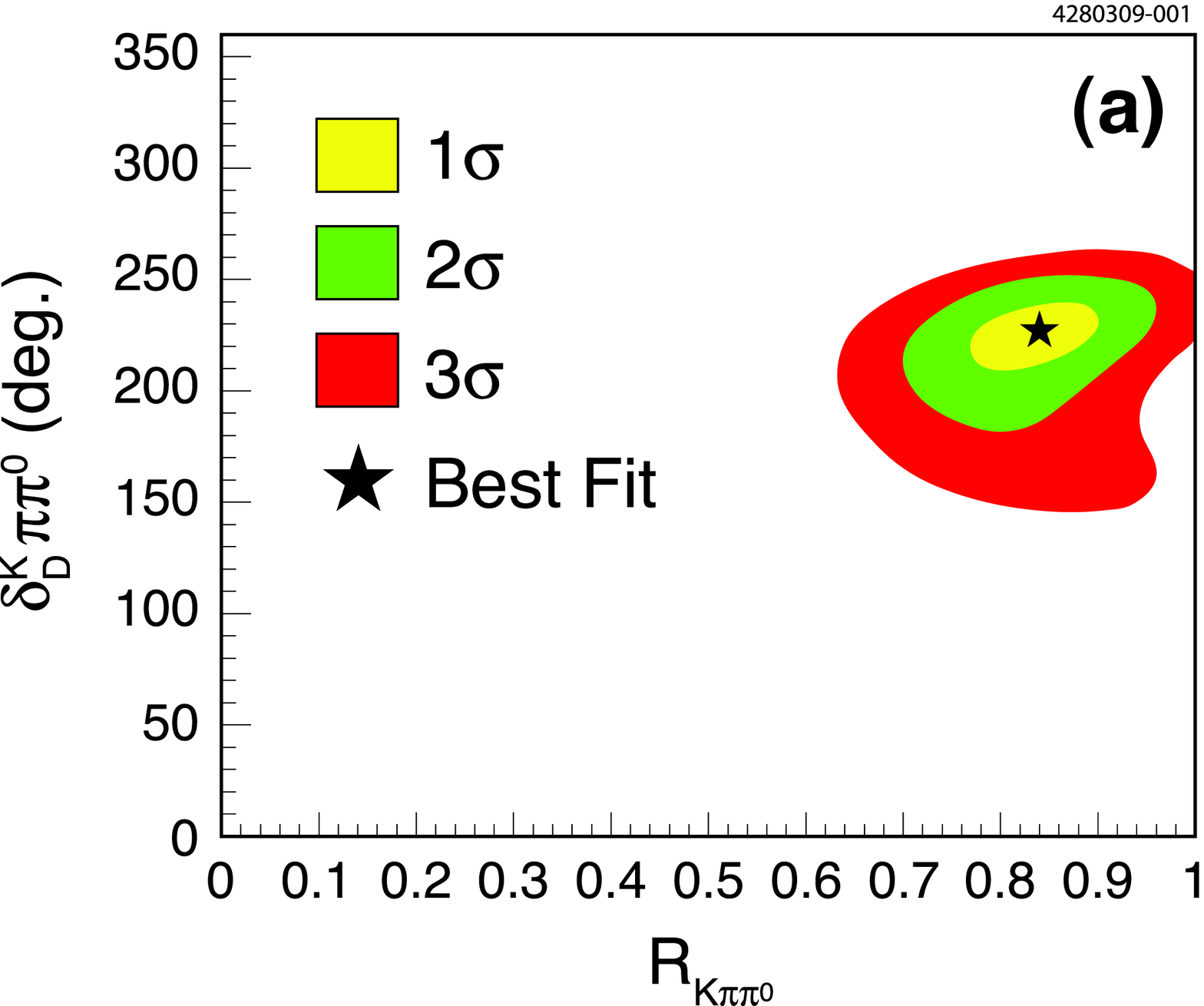} & 
\includegraphics*[width=0.495\columnwidth]{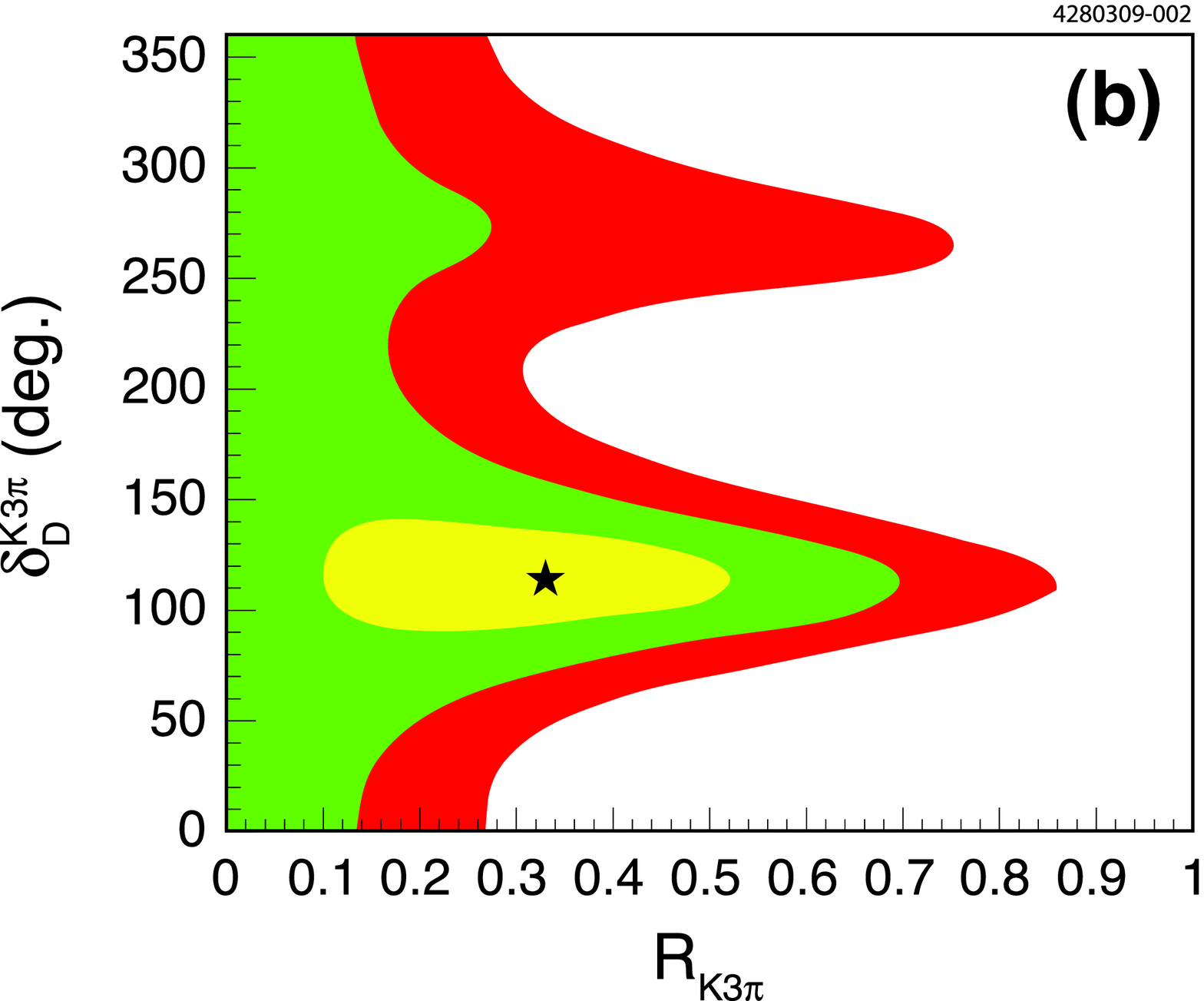}
\end{tabular}
\caption{The $1\sigma$, $2\sigma$ and $3\sigma$ allowed regions of (a) $(R_{K\pi\pi^{0}},\delta_{D}^{K\pi\pi^{0}})$ 
and 
(b)  $(R_{K3\pi},\delta_{D}^{K3\pi})$ parameter space.} \label{fig:ghana}  
\end{figure}

Figures~\ref{fig:ghana}(a) and \ref{fig:ghana}(b) show the $1\sigma$, $2\sigma$ and $3\sigma$ 
allowed regions of $(R_{K\pi\pi^{0}},\delta_{D}^{K\pi\pi^{0}})$ and $(R_{K3\pi},\delta_{D}^{K3\pi})$ parameter space 
from the mixing-constrained fit, respectively. The likelihood is computed as $\mathcal{L}= 
e^{-(\chi^{2}-\chi^{2}_{\mathrm{min}})/2}$ at a point in parameter space; the fit is repeated at each point with the 
values of $R_{F}$ and $\delta_{D}^{F}$ fixed. 
The $95\%$ confidence level (CL) intervals for the parameters are found by integrating one-dimensional likelihood 
scans within the physically allowed region. The following $95\%$ CL intervals are found: $0.70<R_{K\pi\pi^{0}}<0.95$, 
$167^{\circ}<\delta_{D}^{K\pi\pi^{0}}<249^{\circ}$ and $R_{K3\pi}<0.62$. There is no bound on $\delta_{D}^{K3\pi}$ at 
the $95\%$ CL. 


The fit is repeated with the constraints on $x$, $y$ and $\delta_{K\pi}$ removed to estimate these parameters from the 
data; this procedure is referred to as the mixing-unconstrained fit. The $\Delta_{CP}^{F}$ observables are dependent 
on the value of $\delta_{D}^{K\pi}$ and its uncertainty from the normalisation method that used the measured values of 
$S(CP|K^{-}\pi^{+})$. Therefore, initially the value and uncertainties of $\Delta_{CP}^{F}$ are recalculated assuming 
$\cos{\delta_D^{K\pi}}=0\pm 1$ and the mixing-unconstrained fit is performed. The resulting value of 
$\delta_{D}^{K\pi}$ is used to recalculate $\Delta_{CP}^{F}$ and the mixing-unconstrained fit is repeated. This 
procedure is iterated until the parameter values returned by the fit no longer changed within the quoted precision. 
The results of the final iteration are shown in Tab.~\ref{tab:results}. The best-fit values of $x$, $y$, and 
$\delta_D^{K\pi}$ are: $x=(-0.8^{+2.9}_{-2.5})\%$, $y=(-0.7^{+2.4}_{-2.7})\%$, and 
$\delta_{D}^{K\pi}=(-130_{-28}^{+38})$. There is an ambiguity in the solution of the unconstrained-fit if the signs of 
$\delta_{D}^{K\pi}$, $\delta_{D}^{K3\pi}$, $\delta_{D}^{K\pi\pi^{0}}$ and $x$ are all reversed. The correlations 
amongst the fit parameters maybe found in Ref.~\cite{bib:correlations}.  

In summary, the first determination of the coherence factors and average strong-phase differences for $D^{0}\to 
K^{-}\pi^{+}\pi^{0}$ and $D^{0}\to K^{-}3\pi$ has been presented. The results show significant coherence for $D^{0}\to 
K^{-}\pi^{+}\pi^{0}$, but no significant coherence for $D^{0}\to K^{-}3\pi$. These results will improve the 
measurement of the unitarity triangle angle $\gamma$ and the amplitude ratio $r_{B}$ in $B^{-}\to DK^{-}$ decays, 
where the $D$ decays to $K^{-}\pi^{+}\pi^{0}$ and $K^{-}3\pi$. The preliminary result for $R_{K3\pi}$ and 
$\delta_{D}^{K3\pi}$ \cite{bib:franklampard} combined with CLEO-c's measurement of $\delta_{D}^{K\pi}$ 
\cite{bib:tqcaprd} was shown to improve the expected sensitivity to $\gamma$ at LHCb in a combined ADS analysis of 
$K\pi$ and $K3\pi$ final states by up to  $40\%$ \cite{bib:LHCbgamma}. The sensitivity of these data to $y$ and 
$\delta_{D}^{K\pi}$ is also presented.

%


We gratefully acknowledge the effort of the CESR staff
in providing us with excellent luminosity and running conditions.
This work was supported by
the A.P.~Sloan Foundation,
the National Science Foundation,
the U.S. Department of Energy,
the Natural Sciences and Engineering Research Council of Canada, and
the U.K. Science and Technology Facilities Council.

\clearpage
\bf{\large EPAPS addendum}
\begin{table}[h]
\begin{center}
\caption{Correlation matrix for the mixing-constrained fit. Only elements above the diagonal are 
shown.}\label{tab:correlation_constrained}
\begin{tabular}{lcccccccccccc} \hline \hline 
                           & $\delta_{D}^{K3\pi}$ & $R_{K\pi\pi^{0}}$ & $\delta_{D}^{K\pi\pi^{0}}$ & $x$ & $y$ & 
$\delta_{D}^{K\pi}$ 
                           & $\mathcal{B}_{1}$ & $\mathcal{B}_{2}$ 
                           &  $\mathcal{B}_{3}$ & $\mathcal{B}_{4}$ 
                           &  $\mathcal{B}_{5}$ & $\mathcal{B}_{6}$ \\ \hline
$R_{K3\pi}$               & -0.067 & 0.078 & 0.045 & -0.082 & -0.020 & -0.014 &  0.002 &  0.008 &  0.071 &  0.325 & 
-0.134 &  0.051 \\
$\delta_{D}^{K3\pi}$      & ---    & 0.127 & 0.256 & -0.008 &  0.140 &  0.188 & -0.023 &  0.096 &  0.244 & -0.031 & 
-0.126 & -0.032 \\
$R_{K\pi\pi^{0}}$         & ---    & ---   & 0.455 &  0.080 & -0.059 & -0.046 & -0.014 &  0.060 &  0.018 &  0.098 & 
-0.138 &  0.150 \\
$\delta_{D}^{K\pi\pi^{0}}$& ---    & ---   & ---   & -0.033 &  0.377 &  0.467 &  0.004 & -0.027 &  0.142 &  0.131 & 
-0.295 &  0.114 \\
$x$                       & ---    & ---   & ---   & ---    & -0.189 & -0.188 & -0.001 &  0.005 & -0.037 &  0.001 &  
0.047 & -0.006 \\
$y$                       & ---    & ---   & ---   & ---    & ---    &  0.945 &  0.004 & -0.015 &  0.107 & -0.014 & 
-0.146 &  0.012 \\
$\delta_{D}^{K\pi}$       & ---    & ---   & ---   & ---    & ---    & ---    &  0.005 & -0.004 &  0.121 & -0.002 & 
-0.071 &  0.008 \\
$\mathcal{B}_{1}$         & ---    & ---   & ---   & ---    & ---    & ---    & ---    &  0.006 & -0.005 &  0.008 &  
0.001 & -0.002 \\
$\mathcal{B}_{2}$         & ---    & ---   & ---   & ---    & ---    & ---    & ---    & ---    &  0.005 & -0.028 & 
-0.024 &  0.008 \\
$\mathcal{B}_{3}$         & ---    & ---   & ---   & ---    & ---    & ---    & ---    & ---    & ---    &  0.104 &  
0.047 & -0.001 \\
$\mathcal{B}_{4}$         & ---    & ---   & ---   & ---    & ---    & ---    & ---    & ---    & ---    & ---    & 
-0.054 & -0.006 \\
$\mathcal{B}_{5}$         & ---    & ---   & ---   & ---    & ---    & ---    & ---    & ---    & ---    & ---    & 
---    &  0.028 \\ \hline\hline
\multicolumn{13}{l}{Key of branching fractions $(\mathcal{B})$:} \\
\multicolumn{13}{l}{$\mathcal{B}_{1} \equiv \mathcal{B}(D^{0}\to K^{-}\pi^{+})$} \\ 
\multicolumn{13}{l}{$\mathcal{B}_{2} \equiv \mathcal{B}(D^{0}\to K^{+}\pi^{-})$} \\ 
\multicolumn{13}{l}{$\mathcal{B}_{3} \equiv \mathcal{B}(D^{0}\to K^{-}\pi^{+}\pi^{+}\pi^{-})$} \\ 
\multicolumn{13}{l}{$\mathcal{B}_{4} \equiv \mathcal{B}(D^{0}\to K^{+}\pi^{-}\pi^{-}\pi^{+})$} \\ 
\multicolumn{13}{l}{$\mathcal{B}_{5} \equiv \mathcal{B}(D^{0}\to K^{-}\pi^{+}\pi^{0})$} \\ 
\multicolumn{13}{l}{$\mathcal{B}_{6} \equiv \mathcal{B}(D^{0}\to K^{+}\pi^{-}\pi^{0})$}  
\end{tabular}
\end{center}
\end{table}

\begin{table}
\begin{center}
\caption{Correlation matrix for the mixing-unconstrained fit. Only elements above the diagonal are 
shown.}\label{tab:correlation_unconstrained}
\begin{tabular}{lcccccccccccc} \hline \hline 
                           & $\delta_{D}^{K3\pi}$ & $R_{K\pi\pi^{0}}$ & $\delta_{D}^{K\pi\pi^{0}}$ & $x$ & $y$ & 
$\delta_{D}^{K\pi}$ 
                           & $\mathcal{B}_{1}$ & $\mathcal{B}_{2}$ 
                           &  $\mathcal{B}_{3}$ & $\mathcal{B}_{4}$ 
                           &  $\mathcal{B}_{5}$ & $\mathcal{B}_{6}$ \\ \hline
$R_{K3\pi}$                & -0.093 & -0.293 & -0.546 & -0.558 & -0.175 & -0.505 &  0.002 &  0.009 &  0.038 &  0.245 & 
-0.079 &  0.069 \\
$\delta_{D}^{K3\pi}$       & ---    &  0.763 & -0.179 &  0.577 & -0.819 &  0.012 & -0.002 &  0.014 &  0.231 & -0.166 & 
-0.121 & -0.045 \\
$R_{K\pi\pi^{0}}$          & ---    &  ---   & -0.049 &  0.802 & -0.596 & -0.108 & -0.005 &  0.028 &  0.058 & -0.124 & 
-0.006 &  0.029 \\
$\delta_{D}^{K\pi\pi^{0}}$ & ---    &  ---   &  ---   &  0.175 &  0.504 &  0.692 &  0.002 & -0.026 &  0.194 & -0.003 & 
-0.308 &  0.034 \\
$x$                        & ---    &  ---   &  ---   &  ---   & -0.232 &  0.092 &  0.003 & -0.003 &  0.035 & -0.132 &  
0.072 & -0.042 \\
$y$                        & ---    &  ---   &  ---   &  ---   &  ---   &  0.255 & -0.004 &  0.015 & -0.081 &  0.144 &  
0.029 &  0.008 \\
$\delta_{D}^{K\pi}$        & ---    &  ---   &  ---   &  ---   &  ---   &  ---   &  0.006 & -0.034 &  0.211 & -0.157 & 
-0.223 & -0.038 \\
$\mathcal{B}_{1}$          & ---    &  ---   &  ---   &  ---   &  ---   &  ---   &  ---   &  0.005 & -0.003 &  0.006 & 
-0.000 & -0.002 \\
$\mathcal{B}_{2}$          & ---    &  ---   &  ---   &  ---   &  ---   &  ---   &  ---   &  ---   & -0.007 & -0.018 & 
-0.012 &  0.008 \\
$\mathcal{B}_{3}$          & ---    &  ---   &  ---   &  ---   &  ---   &  ---   &  ---   &  ---   &  ---   &  0.075 & 
-0.013 &  0.000 \\
$\mathcal{B}_{4}$          & ---    &  ---   &  ---   &  ---   &  ---   &  ---   &  ---   &  ---   &  ---   &  ---   & 
-0.017 & -0.005 \\
$\mathcal{B}_{5}$          & ---    &  ---   &  ---   &  ---   &  ---   &  ---   &  ---   &  ---   &  ---   &  ---   &  
---   & 0.017  \\ \hline\hline
\end{tabular}
\end{center}
\end{table}


\begin{thebibliography}{99}
\bibitem{bib:conjugation} Here and throughout this letter the charge-conjugate state is implied unless otherwise 
stated.
\end{thebibliography}
\end{document}